\documentclass[12pt]{article}

\usepackage[utf8]{inputenc}
\usepackage{amsmath}
\usepackage{amsfonts}
\usepackage{amssymb}
\usepackage{float}

\def\lb{\label}
\def\be{\begin{equation}}
\def\ee{\end{equation}}

\newcommand{\hC}{\widehat{C}}

\newcommand{\bI}{\mathbf{I}}
\newcommand{\bP}{\mathbf{P}}
\newcommand{\bK}{\mathbf{K}}

\newcommand{\p}[1]{(\ref{#1})}
\newcommand{\bea}{\begin{eqnarray}}
\newcommand{\eea}{\end{eqnarray}}

\newcommand{\ba}{\begin{array}} \newcommand{\ea}{\end{array}}

\DeclareMathOperator{\ad}{ad}
\DeclareMathOperator{\tr}{\bf Tr}

\DeclareMathOperator{\proj}{P}

\begin{document}

\thispagestyle{empty}
\begin{center}
{\huge\bf Split Casimir operator and universal formulation of the simple  Lie algebras}
\end{center}
\vspace{1cm}

\begin{center}
{\Large \bf  A.P.~Isaev${}^{a,b,c}$, S.O.~Krivonos${}^{a}$}
\end{center}

\vspace{0.2cm}

\begin{center}
{${}^a$ \it
Bogoliubov  Laboratory of Theoretical Physics,\\
Joint Institute for Nuclear Research,
141980 Dubna, Russia}\vspace{0.1cm} \\
{${}^b$ \it
St.Petersburg Department of
 Steklov Mathematical Institute of RAS,\\
Fontanka 27, 191023 St. Petersburg,  Russia}\vspace{0.1cm} \\
{${}^c$ \it Faculty of Physics,
Lomonosov Moscow State University, Moscow, Russia}\vspace{0.3cm}

{\tt isaevap@theor.jinr.ru, krivonos@theor.jinr.ru}
\end{center}
\vspace{3cm}

\begin{abstract}\noindent
 We construct characteristic identities for the split (polarized)
  Casimir operators of the simple Lie algebras in  adjoint representation.
  By means of these characteristic identities,   for all simple Lie algebras
  we derive explicit formulae for invariant projectors onto
  irreducible subrepresentations in $T^{\otimes 2}$ in the case when
   $T$ is the  adjoint  representation. These projectors and characteristic identities are considered from the viewpoint of the universal description of
   the simple Lie algebras in terms of the Vogel parameters.
\end{abstract}

\newpage
\setcounter{page}{1}
\section{Introduction}
In this  paper, we demonstrate the usefulness of  the $\mathfrak{g}$-invariant split Casimir operator $\hC$ (see definition in Section {\bf \ref{splkaz}}) in the  representation theory of Lie algebras (see also \cite{Okub}). Namely,  for all simple Lie algebras $\mathfrak{g}$, explicit formulas can be found for invariant projectors onto irreducible representations  that appear in the expansion of the tensor product $T \otimes T'$ of two representations $T$ and $T'$. In particular
these invariant projectors  are constructed  in terms of the 
 matrix $(T \otimes T') \hC$. We stress that
it is natural to find  invariant projectors in terms of $\mathfrak{g}$-invariant operators, which in general  are images of special elements of the so-called  centralizer algebra.

 In the  paper, we consider a very particular problem of constructing
invariant projectors in representation spaces of $T^{\otimes 2}$, where $T \equiv \ad$ is the adjoint representation but for all simple Lie algebras $\mathfrak{g}$. Our approach is closely related to the one outlined in \cite{Okub}, \cite{Cvit}.
In \cite{Cvit}, such invariant projectors were obtained in terms of several special  invariant operators and the calculations were performed using a peculiar diagram technique. In our approach, we try to construct invariant projectors in the representation space $V^{\otimes 2}$ of $T^{\otimes 2}$ by using only one $\mathfrak{g}$-invariant operator which is the split Casimir operator $\hC$.

It turns out (see \cite{IsKri}) that for all simple Lie algebras
$\mathfrak{g}$ in the defining representations  all invariant projectors in $V^{\otimes 2}$ are constructed as polynomials
in $\hC$. It is not the case for the adjoint representation,  i.e. not for all algebras $\mathfrak{g}$ the invariant projectors  in $V_{\ad}^{\otimes 2}$ are constructed as polynomials  of only one operator $\hC_{\ad} \equiv \ad^{\otimes 2}\hC$. Namely, in the case of $s\ell(N)$ and $so(8)$ algebras  there are additional $\mathfrak{g}$-invariant operators which are independent of $\hC_{\ad}$ and act, respectively, in the anti-symmetrized and symmetrized parts of the space $V_{\ad}^{\otimes 2}$. In \cite{IsKri} we construct such additional operators explicitly.

Our study of the split Casimir operator $\hC$ was motivated by the works  \cite{MkSV}, \cite{MkrV}, \cite{MMM}
 and \cite{MMM1}, and by the idea  that the knowledge of the characteristic identities for $\hC_{\ad}$ turns out to be a key point for understanding the so-called universal formulation of the simple Lie algebras \cite{Fog} (see also the historical notes in \cite{Cvit}, section 21.2). Though some characteristic identities  and formulas for certain    $\mathfrak{g}$-invariant projectors can be found in a different form in \cite{Cvit},   we believe that the methods we used and the results obtained can be useful for future research, e.g. from the viewpoint of technical applications  of the split Casimir operator.

The split Casimir operator $\hC_{\ad}$ for the Lie algebras of the classical series in the adjoint representation and for the exceptional algebras  were considered in detail in \cite{IsPr} and \cite{IsKri}.
Here, we present only universal description of these results.

In our paper, to simplify the notation, we everywhere write $s\ell(N)$, $so(N)$  and $sp(2n)$ instead of $s\ell(N,\mathbb{C})$, $so(N,\mathbb{C})$  and $sp(2n,\mathbb{C})$.

\section{Split Casimir operator for simple Lie algebras\label{splkaz}}
\subsection{General definitions}
Let $\mathfrak{g}$ be a simple Lie algebra with  the basis $X_a$ and defining relations
\be\lb{lialg}
[X_a, \; X_b] = C^d_{ab} \; X_d \; ,
\ee
where $C^d_{ab}$ are the structure constants. The Cartan-Killing  metric is defined in the standard way
\be\lb{li04}
 {\sf g}_{ab} \equiv C^{d}_{ac} \, C^{c}_{bd} =   \tr(\ad(X_a)\cdot \ad(X_b)) \; ,
\ee
where $\ad$ denotes adjoint representation: $\ad(X_a)^d_b=C^{d}_{ab}$. Recall that the
 structure constants $C_{abc} \equiv C^{d}_{ab} \, {\sf g}_{dc}$  are antisymmetric under permutation of indices  $(a,b,c)$. We denote an enveloping algebra of the Lie algebra  $\mathfrak{g}$ as ${\cal U}(\mathfrak{g})$.

 Let ${\sf g}^{df}$ be the inverse matrix to the Cartan-Killing  metric (\ref{li04}). We use this matrix and construct the operator
 \be
 \lb{kaz-01}
\hC  = {\sf g}^{ab} X_a \, \otimes \,
  X_b \;\; \in \;\; \mathfrak{g} \, \otimes \,  \mathfrak{g}
   \;\; \subset \;\; {\cal U}(\mathfrak{g})\, \otimes \, {\cal U}(\mathfrak{g}) \; ,
 \ee
which is called the {\it split (or polarized) Casimir operator} of the Lie algebra $\mathfrak{g}$.
This operator is related to the usual quadratic Casimir operator
 \be
 \lb{kaz-c2}
 C_{(2)} = {\sf g}^{ab} \; X_a \cdot X_b \;\; \in \;\;
 {\cal U}(\mathfrak{g}) \; ,
  \ee
 by means of the formula
 \be
 \lb{adCC1}
 \Delta(C_{(2)}) = C_{(2)} \otimes I + I \otimes C_{(2)} + 2 \, \hC  \; ,
 \ee
 where $\Delta$ is the standard co-multiplication
 for enveloping algebras ${\cal U}(\mathfrak{g})$:
 \be
 \lb{mrep2}
 \Delta(X_a) = (X_a \otimes I + I \otimes X_a) \; .
 \ee

The following statement holds (see, for example,
 \cite{Book1}, \cite{ToHa}).
\newtheorem{pro1}{Proposition}[subsection]
\begin{pro1}\label{pro1}
The operator $\hC$, given in (\ref{kaz-01}), does not depend on
the choice of the basis in $\mathfrak{g}$
and satisfies the condition (which is called $\ad$-invariance or
$\mathfrak{g}$-invariance):
 \be
 \lb{kaz-02}
[\Delta(A), \, \hC  ]=
[(A \otimes I + I \otimes A), \, \hC  ]= 0 \; , \;\;\;\;
\forall A \in \mathfrak{g} \; ,
 \ee
 where $\Delta$ is co-multiplication (\ref{mrep2}). In addition, the operator
 $\hC$ obeys the equations
 \be
 \lb{kaz-03}
 [\hC _{12}, \, \hC _{13} + \hC _{23} ] = 0 \;\;\; \Rightarrow \;\;\;
[\hC _{13}, \, \hC _{23} ]
= \frac{1}{2} \; [\hC _{12}, \, \hC _{13} - \hC _{23} ] \; ,
 \ee
which use the standard notation
  \be
 \lb{kaz-01ya}
 \hC _{12} = {\sf g}^{ab} X_a \otimes  X_b \, \otimes  I  , \;
 \hC _{13} = {\sf g}^{ab} X_a  \otimes  I  \otimes   X_b   , \;
 \hC _{23} = {\sf g}^{ab} I  \otimes   X_a  \otimes \,  X_b   .
 \ee
 Here $I$ is the unit element in ${\cal U}(\mathfrak{g})$ and
 $\hC _{ij} \in {\cal U}(\mathfrak{g}) \, \otimes \,
{\cal U}(\mathfrak{g}) \, \otimes \, {\cal U}(\mathfrak{g})$.
\end{pro1}

 \vspace{0.2cm}

Relations (\ref{kaz-03}) indicate that the split Casimir operator (\ref{kaz-01}) realizes the Kono-Drinfeld Lie algebra
  and can be used as a building block for constructing solutions to the quantum and semi-classical Yang-Baxter equations (see e.g. \cite{ChPr}, \cite{Ma} and references therein).

\subsection{The split Casimir operator for simple  Lie algebras in the adjoint representation\label{hCad}}

The generators  $X_a$ of a simple Lie algebra $\mathfrak{g}$ satisfy   the defining relations (\ref{lialg}) and, in the adjoint representation,  $X_a$ are implemented as matrices  ad$(X_a)^d_{\; b} = C^d_{ab}$. In this case
 the split Casimir operator (\ref{kaz-01}) is written as
 \be
 \lb{adCC}
 (\hC_{\ad})^{a_1 a_2}_{\; b_1 b_2} \equiv
 (\ad \otimes \ad)^{a_1 a_2}_{\; b_1 b_2}  (\hC) =
 C^{a_1}_{h b_1} \, C^{a_2}_{f b_2} \, {\sf g}^{h f} \; .
\ee
By definition this operator satisfies identities (\ref{kaz-03}). Below we need one more $\ad$-invariant
rank-1  operator
 \be
 \lb{adK}
 ({\bf K})^{a_1 a_2}_{b_1 b_2} =
 {\sf g}^{a_1 a_2} \; {\sf g}_{b_1 b_2} \; .
\ee
The operators (\ref{adCC}) and (\ref{adK}) act in the tensor product  $V_{\ad} \otimes V_{\ad}$ of two spaces
 $V_{\ad} = \mathfrak{g}$ of the adjoint representation   and have the symmetry properties
 $(\hC_{\ad})^{a_1 a_2}_{\; b_1 b_2} = (\hC_{\ad})^{a_2 a_1}_{\; b_2 b_1}$
 and $\bK^{a_1 a_2}_{\; b_1 b_2} = \bK^{a_2 a_1}_{\; b_2 b_1}$, which are conveniently written in the form
$$
 (\hC_{\ad})_{21} = \bP(\hC_{\ad})_{12}\bP =
  (\hC_{\ad})_{12} \; , \;\;\;
 {\bf K}_{21} = \bP \, {\bf K}_{12} \, \bP = {\bf K}_{12} \; ,
 $$
where $1,2$  are numbers of spaces $V_{\ad}$  in the product $(V_{\ad} \otimes V_{\ad})$ and
 $\bP$ is a permutation matrix in $(V_{\ad} \otimes V_{\ad})$:
 \be
 \lb{perm0}
 \bP (X_{a_1} \otimes X_{a_2}) = (X_{a_2} \otimes X_{a_1}) =
 (X_{b_1} \otimes X_{b_2}) \bP^{b_1 b_2}_{a_1 a_2} \; , \;\;\;
 \bP^{b_1 b_2}_{a_1 a_2} = \delta^{b_1}_{a_2} \, \delta^{b_2}_{a_1} \; .
 \ee
 Here $(X_{a} \otimes X_{b})$ is the basis in the space  $(V_{\ad} \otimes V_{\ad})$. Define the symmetrized and
 anti-symmetrized parts of the operator $\hC_{\ad}$
 \be
 \lb{adCpm}
 (\hC_{\pm})^{a_1 a_2}_{b_1 b_2} =
 \frac{1}{2} ((\hC_{\ad})^{a_1 a_2}_{b_1 b_2} \pm
 (\hC_{\ad})^{a_2 a_1}_{b_1 b_2}) \; ,
 \;\;\;\;\;
  \hC_{\pm} = \bP^{(ad)}_{\pm} \; \hC_{\ad}  =
  \hC_{\ad}\; \bP^{(ad)}_{\pm} \; ,
 \ee
 where $\bP^{(ad)}_{\pm} = \frac{1}{2} (\bI \pm \bP)$ and  $\bI$ is the unit operator in $(V_{\ad})^{\otimes 2}$.
\newtheorem{proCK}[pro1]{Proposition}
\begin{proCK}\lb{proCK}
 The operators $\hC_{\ad}$, $\hC_{\pm}$ and ${\bf K}$, given in  (\ref{adCC}), (\ref{adK})
 and (\ref{adCpm}), satisfy the identities
 \be
 \lb{idCC}
 \hC_{-}^2 = - \frac{1}{2}  \hC_{-}  \; ,
 \ee
  \be
 \lb{idK}
 \hC_{-} \, {\bf K} = 0 = {\bf K} \, \hC_{-}  \; , \;\;\;
 \hC_{\ad} \, {\bf K} = {\bf K} \, \hC_{\ad} = - {\bf K} \; ,
 \ee
 \be
 \lb{idK2}
 \hC_{+} \, {\bf K} = {\bf K} \, \hC_{+} = - {\bf K} \; .
 \ee
\end{proCK}

The proof was presented in \cite{IsKri}.

Now we take into account definitions (\ref{adCC}), (\ref{adK}), (\ref{perm0}), relations (\ref{idCC}),
\be
 \lb{idkm}
 (\hC_{-})^{a_1 a_2}_{b_1 b_2} = - \frac{1}{2}
  \, C^{a_1 a_2}_d \; C^d_{b_1 b_2} \; , \;\;\;\;\;
 C^{a_1 a_2}_d  \equiv C^{a_1}_{d \, b_2} \; {\sf g}^{b_2 a_2} \; ,
 \ee
 \be
 \lb{casad}
 C^d_{b_1 b_2} C^{b_1 b_2}_a = \delta^d_a \;\;\;\;\;\;\;
 \Leftrightarrow \;\;\;\;\;\;\;
  {\rm ad}(C_{(2)})^f_{\;\; r} =
 {\sf g}^{ab} \, C^{f}_{a\, d} C^{d}_{b\, r} = \delta^f_r\;,
 \ee
  and  $C_{ba}^a = 0$, which is valid for all simple Lie algebras, and obtain general formulas for the traces
 \be
 \lb{trac1}
 \begin{array}{c}
 {\bf Tr}(\hC_{\ad}) = 0\,  , \;\;\;
 {\bf Tr}(\hC_{\pm}) = \pm \frac{1}{2} \dim \mathfrak{g} \, , \;\;\;
 {\bf Tr} (\hC_{\ad}^2) = \dim \mathfrak{g} \, , \\ [0.3cm]
 {\bf Tr}(\hC_{-}^2) = - \frac{1}{2}  {\bf Tr}(\hC_{-}) =
   \frac{1}{4}  \dim \mathfrak{g} , \\ [0.3cm]
   {\bf Tr}(\hC_{+}^2) = {\bf Tr}(\hC_{\ad}^2 - \hC_{-}^2) =
   \frac{3}{4}  \dim \mathfrak{g} , \\ [0.3cm]
    {\bf Tr}(\bK) = \dim \mathfrak{g} \, , \;\;\;
    {\bf Tr}(\bI ) = (\dim \mathfrak{g})^2 \, , \;\;\;
      {\bf Tr}(\bP) = \dim \mathfrak{g} \; .
 \end{array}
 \ee
 where ${\bf Tr} \equiv {\rm Tr}_{1}{\rm Tr}_{2}$ is the trace  in the space $V_{\ad} \otimes V_{\ad}$
 (as usual the indices $1$ and $2$ are attributed to factors in the product  $V_{\ad} \otimes V_{\ad}$). These formulas will be
  used in what follows.

Using the characteristic identity (\ref{idCC}) for the operator $\hC _{-}$, one can construct two mutually orthogonal projectors
\be
 \lb{XX12}
 \proj_1 = - 2 \, \hC_{-} \; , \;\;\;\;
 \proj_2 =  2 \, \hC_{-} + \bP^{(\ad)}_{-} \;\;\;\;\; \Rightarrow \;\;\;\;\;
  \proj_i \proj_k = \proj_i \; \delta_{ik} \; ,
 \ee
which decompose the anti-symmetrized  part $\bP^{(ad)}_{-} (\ad \otimes \ad)$
of the representation $(\ad \otimes \ad)$ into two sub-representations ${\sf X}_{1,2} = \proj_{1,2}  (\ad \otimes \ad)$.
Dimensions of these sub-representations are equal to the  traces
  of corresponding projectors (\ref{XX12})
 \be
 \lb{XX123}
 \dim {\sf X}_1 = {\bf Tr}(\proj_1) =
 \dim \mathfrak{g}  , \;
 \dim {\sf X}_2 = {\bf Tr}(\proj_2) =
 \frac{1}{2} \dim \mathfrak{g}\;  (\dim \mathfrak{g} - 3) ,
 \ee
 where we use the general formulae (\ref{trac1}). Since the constants  $C^d_{b_1 b_2}$ play the role of the Clebsch-Gordan coefficients for the  fusion $\ad^{\otimes 2} \to \ad$, we see from the explicit form
 (\ref{idkm}) of the operator $\hC_{-}$ that the projector $\proj_1$, given in (\ref{XX12}), extracts  the adjoint representation  ${\sf X}_1 = \ad$ in $\bP^{(ad)}_{-} (\ad^{\otimes 2})$. Thus, the adjoint representation
 is always contained in  the anti-symmetrized part $\bP^{(ad)}_{-} (\ad^{\otimes 2})$.
 The first formula in (\ref{XX123}) confirms  the  equivalence of ${\sf X}_1$ and $\ad$.
Note also that ${\sf X}_2$   is not necessarily irreducible representation for all simple Lie algebras
(for the details, see \cite{IsKri}).

 \subsection{Universal characteristic identities for operator $\hC_+$
 in the case of Lie algebras of classical series\label{uncha}}

It was shown in \cite{IsKri} that for the  algebras of the classical series  $A_n,B_n,C_n,D_n$
the characteristic identities
 for the operator $\hC_+$ in the adjoint representation can be written in a generic form
   \be
\lb{chcp4}
\hC_+^3 +\frac{1}{2} \hC_+^2 = \mu_1 \hC_+
 + \mu_2 (\bI^{(ad)} + \bP^{(ad)} -2 \bK) \; ,
\ee
where $\mu_1$ and $\mu_2$ are the parameters of the simple Lie algebras ad we define these parameters  at the moment.
Multiplying both sides of equation \p{chcp4} by $\bK$
and using  the relations
$$
\bK\, (\bI^{(ad)} + \bP^{(ad)})= 2 \, \bK \; , \;\;\;\;
\bK \, \hC_{+} = - \bK \; , \;\;\;\;
\bK \cdot \bK = \dim \mathfrak{g} \cdot \bK \; ,
$$
one may express the dimension of the Lie algebra  $\mathfrak{g}$ through the parameters $\mu_1$ and $\mu_2$
 \be
\lb{abcd01}
\dim \mathfrak{g} = \frac{2 \mu_2 - \mu_1 + 1/2}{2 \mu_2}  \; .
\ee
Then, we multiply both sides of \p{chcp4} by $\bP^{(ad)}_+ (\hC_+ + 1)$
and deduce
the characteristic identity for $\hC_+$ projected onto the subspace $\bP^{(ad)}_+(V_{\ad}^{\otimes 2})
\equiv \frac{1}{2}(\bI^{(\ad)} + \bP^{(\ad)})\, (V_{\ad}^{\otimes 2})$:
  \be
\lb{abcd02}
\bP^{(ad)}_+ (\hC_+ + 1) (\hC_+^3 + \frac{1}{2}\hC_+^2
- \mu_1 \hC_+ - 2 \mu_2) = 0  \; ,
\ee
which can be written in a factorized form
 \be
\lb{abcd03}
\bP^{(ad)}_+  (\hC_+ + 1) (\hC_+ + \frac{\alpha}{2t})
 (\hC_+ + \frac{\beta}{2t})(\hC_+ + \frac{\gamma}{2t}) = 0
 \;\; \Leftrightarrow \;\;
 \bP^{(ad)}_+ \prod_{i=1}^4 (\hC_+ - a_i) = 0 \; .
\ee
 Here we introduce the notation for the roots
 of the identity (\ref{abcd02})
 \be
 \lb{root01}
 a_1 = -1 \; , \;\;\;
 a_2 = -\frac{\alpha}{2t} \; , \;\;\;
 a_3 = -\frac{\beta}{2t} \; , \;\;\;  a_4 = -\frac{\gamma}{2t} \; , \quad
  t=\alpha+\beta+\gamma \, ,
 \ee
 and the last equation follows from the condition
 $(a_2+a_3+a_4) = -1/2$ which is obtained from the comparing of (\ref{abcd02}) and (\ref{abcd03}).
 The parameter $t$ normalizes
 the eigenvalues of the operator $\hC_+$.
 For each simple Lie algebra $\mathfrak{g}$ we
 choose $t^{-1}$ such that
 \be\label{t}
 \left( \theta, \theta\right)=\frac{1}{t} \; ,
 \ee
 where  $\theta$ is the highest   root  of $\mathfrak{g}$.
Thus, $t$ coincides with the dual Coxeter number $h^{\vee}$  of the algebra $\mathfrak{g}$.
The parameters $\alpha, \beta,\gamma$ were introduced by Vogel \cite{Fog}.
The values of these parameters for the algebras $A_n,B_n,C_n,D_n$
 are  summarized  in the  Table 1.
  \begin{center}
Table 1. \\\vspace{0.5cm}
\begin{tabular}{|c|c|c|c|c|}
\hline
$\;\;$ & $s\ell(n+1)$ & $so(2n+1)$ & $sp(2n)$ & $so(2n)$ \\
\hline
$t$  &\footnotesize  $n+1$ &
\footnotesize  $2n-1$ &
\footnotesize $n+1$ &\footnotesize  $2n-2$   \\
\hline
$\frac{\alpha}{2 t}$  &\footnotesize  $-1/(n+1)$ &
\footnotesize  $-1/(2n-1)$ &
\footnotesize $-1/(n+1)$ &\footnotesize  $-1/(2n-2)$   \\
\hline
$\frac{\beta}{2 t}$  &\footnotesize $1/(n+1)$ &
\footnotesize $2/(2n-1)$ &
\footnotesize $1/(2n+2)$ &\footnotesize $1/(n-1)$ \\
\hline
$\frac{\gamma}{2 t}$  &\footnotesize $1/2$ &
\footnotesize $(2n-3)/(4n-2)$ &
\footnotesize $(n+2)/(2n+2)$ &\footnotesize $(n-2)/(2 n-2)$ \\
\hline
\end{tabular}
\end{center}
Here we encounter an interesting non-linear Diophantine problem of finding all
integer $\dim \mathfrak{g}$ in (\ref{abcd06})  for which the parameters
$\alpha,\beta$,$\gamma$ and $\dim V_{(a_i)}$ are integers. The partial
solutions of this problem are given in Table 1.
The analogous Diophantine problems were considered in
\cite{Rub}, \cite{RuHu}.

\noindent
Comparison of  equations (\ref{abcd02})
and (\ref{abcd03}) implies that
the parameters $\mu_1$ and $\mu_2$ are
expressed via the Vogel parameters as
 \be
\lb{abcd04}
\mu_1 = - \frac{\alpha\beta  + \alpha\gamma  + \beta\gamma}{4t^2}
 \; , \quad
\mu_2 = - \frac{\alpha\beta \gamma}{16 t^3} \; ,
\ee
and the dimensions \p{abcd01} of the simple Lie algebras
 acquire a remarkable universal form obtained
by Deligne and Vogel \cite{Delig},\cite{Fog}:
  \be
\lb{abcd06}
 \dim \mathfrak{g} =
 \frac{(\alpha-2t)(\beta-2t) (\gamma-2t)}{\alpha\beta \gamma} \; .
\ee
 Now by using the characteristic identity (\ref{abcd03}), one
 can obtain
the universal form of the projectors $\proj_{(a_i)}^{(+)}$ on the invariant
subspaces $V_{(a_i)}$ in the symmetrized space
$\bP^{(\ad)}_+ \, (V_{\ad}^{\otimes 2})$:
$$
 \begin{array}{l}
\proj^{(+)}_{(-\frac{\alpha}{2t})} = 
 \frac{4t^2}{(\beta-\alpha)(\gamma-\alpha)}
 \Bigl(  \hC_+^2 + \bigl(\frac{1}{2} -\frac{\alpha}{2t}\bigr) \hC_+
 + \frac{\beta \gamma}{8 t^2}
 \bigl( \bI^{(\ad)} +  \bP^{(\ad)}
 - \frac{2 \alpha}{(\alpha - 2t)}\bK \bigr) \, \Bigr)
 \equiv \proj^{(+)}(\alpha|\beta,\gamma) \; ,
 \end{array}
$$
$$
\proj^{(+)}_{(-\frac{\beta}{2t})} = \proj^{(+)}(\beta|\alpha,\gamma)
\; , \;\;\;\;\;\;
\proj^{(+)}_{(-\frac{\gamma}{2t})} = \proj^{(+)}(\gamma|\alpha,\beta)
\; , \;\;\;\;\;\;
\proj^{(+)}_{(-1)} = \frac{1}{\dim \mathfrak{g}}\;  \bK \; .
$$
The irreducible representations that act in the subspaces
 $V_{(-1)}$, $V_{(-\frac{\alpha}{2t})}$,
 $V_{(-\frac{\beta}{2t})}$, $V_{(-\frac{\gamma}{2t})}$
 were respectively denoted in \cite{Fog} as
 ${\sf X}_0$, $Y_2(\alpha)$, $Y_2(\beta)$, $Y_2(\gamma)$;
 see Section {\bf \ref{Vogel}} below.
Finally, we calculate (by means of trace formulas (\ref{trac1}))
 the universal expressions \cite{Fog}
 for the dimensions of the invariant eigenspaces $V_{(a_i)}$:
  \bea
  &&\dim V_{(-1)} ={\bf Tr} \, \proj^{(+)}_{(-1)} = 1\, , \nonumber \\
  &&\dim V_{(-\frac{\alpha}{2t})} =
 {\bf Tr} \, \proj^{(+)}_{(-\frac{\alpha}{2t})}
 =-\frac{(3\alpha-2t)(\beta-2t)(\gamma-2t)t(\beta+t)(\gamma+t)}{
\alpha^2(\alpha-\beta)\beta(\alpha-\gamma)\gamma} \, , \lb{unidim02a}\\
 \lb{unidim02b}
  &&\dim V_{(-\frac{\beta}{2t})} =
 {\bf Tr} \, \proj^{(+)}_{(-\frac{\beta}{2t})}
 =-\frac{(3\beta-2t)(\alpha-2t)(\gamma-2t)t(\alpha+t)(\gamma+t)}{
\beta^2(\beta-\alpha)\alpha(\beta-\gamma)\gamma} \, ,\\
&& \dim V_{(-\frac{\gamma}{2t})} =
{\bf Tr} \, \proj^{(+)}_{(-\frac{\gamma}{2t})}
 = -\frac{(3\gamma-2t)(\beta-2t)(\alpha-2t)t(\beta+t)(\alpha+t)}{
\gamma^2(\gamma-\beta)\beta(\gamma-\alpha)\alpha}  \, . \lb{unidim02c}
\eea

\subsection{Universal characteristic identities for  operator $\hC$ in the case of  exceptional Lie algebras
\label{uniex}}

The antisymmetric parts of the split Casimir operators
$\hC_-$ for all simple Lie algebras in the adjoint representation  obey the same identity (\ref{idCC})
\be\label{idCCa}
\hC_- \left( \hC_-+\frac{1}{2}\right) =0 .
\ee
The symmetric parts of the split Casimir operators $\hC_+$ in the adjoint representation
for the exceptional Lie algebras  obey  identities  which have a
similar structure\footnote{The universal
 formulae (\ref{genid}) was obtained  in \cite{Cvit}, eq. (17.10),
under the assumption that $\hC_+^{\, 2}$ is expressed as
 a linear combination of $\mathfrak{g}$-invariant
 operators $(\bI+\bP)$, $\bK$ and $\hC_+$.
 We explicitly checked this assumption
 for all exeptional Lie algebras \cite{IsKri}.}
\be
\lb{genid}
\hC_+^2 = - \frac{1}{6} \hC_+ + \mu \; \left( \bI + \bP +\bK\right) \; ,
\ee
where the universal parameter  $\mu$ is fixed as follows:
 \be
\lb{unimu}
\mu = \frac{5}{6(2 + \dim(\mathfrak{g}))} \; .
\ee
Note that  identities   for the algebras  $s\ell(3)$ and
  $so(8)$ have the same structure.

From \p{genid} one can obtain the universal characteristic identity on the symmetric part of the split Casimir operator $\hC_+$ projected onto the subspace $\bP^{(ad)}_+ (V_{ad}^{(\otimes 2})$
\be
 \lb{chplu}
 \bP^{(ad)}_+(\hC_+ +1)(\hC_+^2 + \frac{1}{6}\hC_+ - 2\mu ) \equiv
 \bP^{(ad)}_+(\hC_+ +1)(\hC_+ + \frac{\alpha}{2 t})
 (\hC_+ + \frac{\beta}{2 t} ) = 0 \; ,
 \ee
where we introduced the notation for two eigenvalues of the $\hC_+$ :
 \be
 \lb{albe}
 \frac{\alpha}{2 t} = \frac{1 - \mu'}{12} \; , \;\;\;\;
 \frac{\beta}{2 t} = \frac{1 + \mu'}{12} \; , \;\;\;\;\;\;
  \mu' := \sqrt{1+288\mu} =
  \sqrt{\frac{\dim \mathfrak{g}+ 242}{\dim \mathfrak{g} + 2}} \; .
 \ee
According to (\ref{chplu}), these parameters are 
related as 
$$
3(\alpha + \beta) = t \; .
$$
With the fixed value of the parameter $\alpha$, this relation defines the line of the exceptional Lie algebras on the $\beta,t$ plane (see eq.(\ref{exline}) below).
Following  \cite{Cvit}, note that $\mu'$ is a rational number only
for a certain sequence of dimensions $\dim \mathfrak{g}$.
It turns out that this sequence is finite\footnote{We thank D.O.Orlov
who proved the finiteness of this sequence.}:
 \be
 \lb{diof1}
\begin{array}{c}
\dim \mathfrak{g} = 3,8,14,28,
47, 52, 78, 96, 119, 133, 190, 248, 287, 336, \\
 484, 603, 782, 1081, 1680, 3479 \; ,
\end{array}
 \ee
which includes the dimensions $14,
52, 78, 133, 248$  of the exceptional Lie algebras
$\mathfrak{g}_2,\mathfrak{f}_4,\mathfrak{e}_6,
 \mathfrak{e}_7, \mathfrak{e}_8$, and the dimensions
 $8$ and  $28$ of the algebras $\mathfrak{s\ell}(3)$ and 
 $\mathfrak{so}(8)$, which are sometimes also
 referred to as exceptional. Thus, for these algebras,
 using (\ref{albe}), we calculate the  values of the parameters
 $\frac{\alpha}{2 t},\frac{\beta}{2 t}$ given in Table 2.
 \begin{center}
Table 2. \\\vspace{0.5cm}
\begin{tabular}{|c|c|c|c|c|c|c|c|}
\hline
$\;\;$ & $s\ell(3)$ & $so(8)$ & $\mathfrak{g}_2$ & $\mathfrak{f}_4$
& $\mathfrak{e}_6$ &
$\mathfrak{e}_7$ & $\mathfrak{e}_8$  \\
\hline
$\frac{\alpha}{2 t}$  &\footnotesize  $-1/3$ &\footnotesize  $-1/6$ &
\footnotesize $-1/4$ &\footnotesize  $-1/9$ &\footnotesize  $-1/12$
&\footnotesize  $-1/18$ &\footnotesize  $-1/30$  \\
\hline
$\frac{\beta}{2 t}$  &\footnotesize $1/2$ &\footnotesize $1/3$ &
\footnotesize $5/12$ &\footnotesize $5/18$ &\footnotesize $1/4$ &
\footnotesize $2/9$ &\footnotesize $1/5$ \\
\hline
\end{tabular}
\end{center}
 Taking into account
 that $\hC_-$  satisfies (\ref{idCC}) and
 $\hC_+$ satisfies (\ref{chplu}),
 we obtain the following identities for the total
 split Casimir operator
 $\hC_{\ad}=(\hC_{+}+ \hC_{-})$ in the case of the exceptional
 Lie algebras:
 \be
 \lb{chexE}
\hC_{\ad} \left(\hC_{\ad}+\frac{1}{2}\right)
\left(\hC_{\ad}+ 1 \right) \left(\hC_{\ad}^2
+ \frac{1}{6} \hC_{\ad}-2  \mu \right) =0 \;\;\; \Rightarrow
\ee
\be
 \lb{chexe}
\hC_{\ad} \left(\hC_{\ad}+\frac{1}{2}\right)
\left(\hC_{\ad}+ 1 \right) \left(\hC_{\ad}+\frac{\alpha}{2t} \right)
\left( \hC_{\ad}+ \frac{\beta}{2t} \right) =0 \; .
\ee
Here $\mu$ is defined in (\ref{unimu}) and
$\frac{\alpha}{2t},\frac{\beta}{2t}$ are given in Table 2.

\noindent
{\bf Remark.} The sequence (\ref{diof1}) contains dimensions
  $\dim \mathfrak{g}^* =(10 m -122 +360/m)$,
 $(m \in \mathbb{N})$
 referring to the adjoint representations of the so-called
 $E_8$ family of algebras $\mathfrak{g}^*$;
see \cite{Cvit}, eq. (21.1). For these dimensions we have the relation
$\mu'=|(m+6)/(m-6)|$. Two numbers $47$ and $119$ from the
sequence (\ref{diof1}) do not belong to the sequence
$\dim \mathfrak{g}^*$. Thus, the
interpretation of these two numbers as  dimensions of some algebras is missing. Moreover, for values $\dim \mathfrak{g}$
 given in (\ref{diof1}), using (\ref{albe}),
 one can calculate dimensions (\ref{unidim02a})
 of the corresponding  representations $Y(\alpha)$:
 $$
 \begin{array}{c}
 \dim V_{(-\frac{\alpha}{2t})} =
 \left\{5,27,77,300,\frac{14553}{17},1053,2430,\frac{48608}{13},
 \frac{111078}{19},7371,15504,27000, \right .
\\ [0.2cm]
\left. \frac{841279}{23},\frac{862407}{17},
107892,\frac{2205225}{13},
\frac{578151}{2},559911,\frac{42507504}{31},
\frac{363823677}{61}\right\} \; .
    \end{array}  
 $$
 Since
 $\dim  V_{(-\frac{\alpha}{2t})}$ should be integer,
 we conclude that no Lie algebras exist with
 dimensions $47,96,119,287,336,
 603,782,1680,3479$, for which we assume characteristic
 identity (\ref{chplu}) and the trace formulas (\ref{trac1}).
 Note (see, e.g., \cite{IsKri}) that 
 the list of $\dim V_{(-\frac{\alpha}{2t})}$
 contains integer dimensions $27,77,300,1053,2430,7371,27000$ 
 of representations (arising in the decomposition
 of $\bP^{(ad)}_+(\ad^{\otimes 2})$) 
 of the corresponding algebras $\mathfrak{s\ell}(3),\mathfrak{g}_2,\mathfrak{so}(8),
 \mathfrak{f}_4,\mathfrak{e}_6,
 \mathfrak{e}_7,\mathfrak{e}_8$.

\section{Universal characteristic identities for
operator $\hC$ and Vogel parameters\label{Vogel}}
In the  sections 2.2 and 2.3 we constructed  the projectors onto the spaces of irreducible sub-representations in the
representation $\ad^{\otimes 2}$ for all
simple Lie algebras of classical series $A_n , B_n ,  C_n , D_n$. In all cases  the construction was
carried out by finding the characteristic identities for the split Casimir operators. We note that the construction of projectors in terms of the split Casimir operator and finding  the dimensions of the corresponding subspaces can be obtained  by using the Vogel parameters $\alpha, \beta$ and $ \gamma $,
which were introduced in \cite{Fog} (see also \cite{Lan, MkrV}).
The values of the Vogel parameters specify simple Lie algebras and we present these values  in the Table 3 (see below). Since all universal formulas for the simple Lie algebras
are written as homogeneous and symmetric functions of the parameters $\alpha, \beta$ and $\gamma $,
one can consider simple Lie algebras as points in the space $\mathbb{RP}^3/\mathbb{S}_3$.
 It is convenient to choose  a normalization  of the parameters
such that $\alpha = -2 $, see Table 3.
Note that the data in the first six lines of Table 3 coincide
with the data given in Table 1 of Section {\bf \ref{uncha}}.
We list the Vogel parameters for the algebras
$s\ell(3)$ and $so(8)$ in the separate lines of Table 5,
 since the characteristic identities for
the symmetric part $\hC_{+}$ of the split
Casimir operator in the adjoint representations have the same order
and the same structure as for the exceptional Lie algebras
(see \cite{IsKri}).

\begin{center}
Table 3. \\ [0.2cm]
\begin{tabular}{|c|c|c|c|c|c|c|c|c|}
\hline
Type & Lie algebra & $\alpha$ & $\beta$ & $\gamma$ & $t$ &
$-\frac{\alpha}{2t}=\frac{1}{t}$ & $-\frac{\beta}{2t}$
& $-\frac{\gamma}{2t}$ \\
\hline
$A_n$ & $s\ell(n+1)$ & $-2$ & $2$ & $n+1$ & $n+1$ & $\frac{1}{n+1}$
& $-\frac{1}{n+1}$ &\footnotesize  $-1/2$ \\
\hline
$B_n$ & $so(2n+1)$ & $-2$ & $4$ & $2n-3$ & $2n-1$ & $\frac{1}{2n-1}$
& $-\frac{2}{2n-1}$ &\footnotesize  $-\frac{2n-3}{2(2n-1)}$ \\
\hline
$C_n$ & $sp(2n)$ & $-2$ & $1$ & $n+2$ & $n+1$ & $\frac{1}{n+1}$
& $-\frac{1}{2(n+1)}$ &\footnotesize  $-\frac{n+2}{2(n+1)}$ \\
\hline
$D_n$ & $so(2n)$ & $-2$ & $4$ & $2n-4$ & $2n-2$ & $\frac{1}{2n-2}$
& $-\frac{1}{n-1}$ &\footnotesize  $-\frac{n-2}{2(n-1)}$ \\
\hline
$A_2$ & $s\ell(3)$ & $-2$ & $2$ & $3$ & $3$ &\footnotesize  $1/3$
&\footnotesize  $-1/3$ &\footnotesize  $-1/2$ \\
\hline
$D_4$ & $so(8)$ & $-2$ & $4$ & $4$ & $6$ & \footnotesize $1/6$
&\footnotesize $-1/3$ &\footnotesize  $-1/3$ \\
\hline
$G_2$ & $\mathfrak{g}_2$ & $-2$ & $10/3$ & $8/3$ & $4$ &\footnotesize $1/4$
&\footnotesize $-5/12$ &\footnotesize  $-1/3$ \\
\hline
$F_4$ & $\mathfrak{f}_4$ & $-2$ & $5$ & $6$ & $9$ &\footnotesize $1/9$
&\footnotesize $ -5/18$ &\footnotesize  $-1/3$  \\
\hline
$E_6$ & $\mathfrak{e}_6$ & $-2$ & $6$ & $8$ & $12$ &\footnotesize $1/12$
&\footnotesize  $-1/4$&\footnotesize  $-1/3$ \\
\hline
$E_7$ & $\mathfrak{e}_7$ & $-2$ & $8$ & $12$ & $18$ &\footnotesize $1/18$
&\footnotesize  $-2/9$ &\footnotesize  $-1/3$ \\
\hline
$E_8$ & $\mathfrak{e}_8$ & $-2$ & $12$ & $20$ & $30$ &\footnotesize $1/30$
&\footnotesize  $-1/5$ &\footnotesize  $-1/3$ \\
\hline
\end{tabular}
\end{center}

As usual, we split the tensor product of two adjoint representations into the symmetric and antisymmetric parts
\be\label{new1}
\ad \otimes \ad = \bP^{(ad)}_+(\ad \otimes \ad)+
\bP^{(ad)}_-(\ad \otimes \ad).
\ee
In the general case of the Lie algebras of the classical
series\footnote{The algebras $s\ell(3)$ and $so(8)$ are exceptional
cases.},
the symmetric part $\bP^{(ad)}_+(\ad^{\otimes 2})$ decomposes
into 4 irreducible representations (see e.g. \cite{Fog}):
a singlet, denoted as ${\sf X}_0$, with zero eigenvalue of the quadratic
Casimir operator $C_{(2)}$ (which corresponds to the   eigenvalue $(-1)$ for the split operator $\hC$),
and 3 representations which we denote as $Y_2(\alpha), Y_2(\beta), Y_2(\gamma)$.
Their dimensions, as well as the corresponding values   of the quadratic  Casimir operator $C_{(2)}$ (defined in (\ref{kaz-c2})) and  split Casimir operator $\hC$ are equal to:
\begin{align}
\dim Y_2(\alpha) & = \dim V_{(-\frac{\alpha}{2t})} \, , \;\;\;
 C_{(2)} =2-\frac{\alpha}{t} \, , \;\;\;
 \hC =-\frac{\alpha}{2t} \, , \\
\dim Y_2(\beta)& =\dim V_{(-\frac{\beta}{2t})}  \, , \;\;\;
 C_{(2)}=2-\frac{\beta}{t}\, , \;\;\;
\hC=-\frac{\beta}{2t} \, ,
\label{dimy2b} \\
\dim Y_2(\gamma)&=\dim V_{(-\frac{\gamma}{2t})}   \, , \;\;\;
C_{(2)}=2-\frac{\gamma}{t} \, , \;\;\;
\hC=-\frac{\gamma}{2t} \, .
\label{dimy2g}
\end{align}
where the explicit expressions for
   $\dim V_{(-\frac{\alpha}{2t})}$,
$\dim V_{(-\frac{\beta}{2t})}$, $\dim V_{(-\frac{\gamma}{2t})}$
 are given in (\ref{unidim02a})--(\ref{unidim02c})
 and the eigenvalues  of the operators $C_{(2)}$ and $\hC$ are related by the condition:
\begin{equation}
\hC
=\frac{1}{2}C_{(2)}-1 \; .
\end{equation}
The eigenvalues of the operator
 $\hC$ on the representations $Y_2(\alpha),Y_2(\beta),Y_2(\gamma)$ in $\bP^{(ad)}_+(\ad\times \ad)$ are presented in
  three last columns of  Table 3. Therefore, taking into account that $\hC_{+}$ has four
 eigenvalues
 $(-1,-\frac{\alpha}{2t},-\frac{\beta}{2t},-\frac{\gamma}{2t})$
 and $\hC_{-}$ has two eigenvalues $(0,-\frac{1}{2})$, the generic
 characteristic identity for the split Casimir operator reads
   \be
 \lb{chvog}
 \hC_{\ad} (\hC_{\ad} + \frac{1}{2})(\hC_{\ad} + 1)
 (\hC_{\ad} + \frac{\alpha}{2t})(\hC_{\ad} + \frac{\beta}{2t})
 (\hC_{\ad} + \frac{\gamma}{2t}) = 0 \; .
 \ee
  In the case of the $s\ell(N)$ algebras, the  eigenvalue $(-1/2)$ of the operator
 $\hC_{\ad}$  is doubly degenerated, since
 $ \frac{\gamma}{2t} = 1/2 $; therefore, in the identity (\ref{chvog})
one should keep only one factor $ (\hC_{\ad} + \frac{1}{2})$ of two.

We now turn to the discussion of the case of the exceptional Lie algebras.
Note that all exceptional Lie algebras are distinguished  in Table 3 by
the value of the parameter $\gamma/(2t)$ equals to $1/3 $ (all other parameters
of the exceptional Lie algebras in Table 3 are in agreement with the parameters listed in Table 2
of Section {\bf \ref{uniex}}).
Thus, all exceptional Lie algebras in the three-dimensional space of the Vogel parameters
$ (\alpha, \beta, \gamma) $ lie  in the plane $\alpha = -2$
 on the line:
\be
 \lb{exline}
 3 \gamma = 2 t \;\;\;\;\; \Rightarrow \;\;\;\;\;
 \gamma = 2 \beta - 4 \; .
 \ee
We chose the coordinates $(\beta, \gamma)$ on this plane  and visualize all simple Lie algebras as points
on this plane (Vogel map).

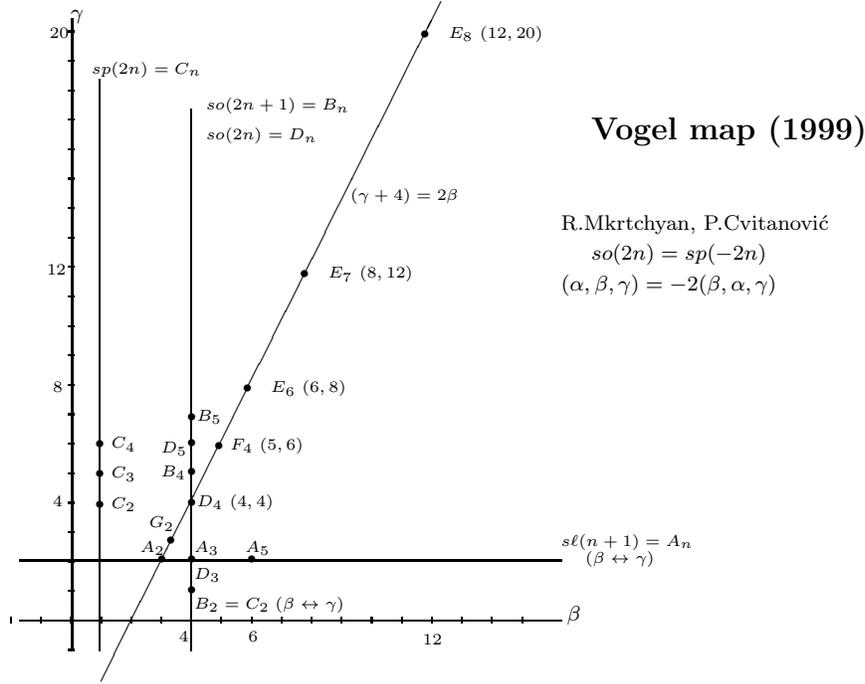
\unitlength=4mm
\begin{picture}(20,21)(-6,0)

{\put(20,16){\bf Vogel map (1999)}}

{\put(19,13){\scriptsize R.Mkrtchyan, P.Cvitanovi\'{c}}}
{\put(20,12){\scriptsize  $so(2n)=sp(-2n)$}}
{\put(19,11){\scriptsize
$(\alpha,\beta,\gamma)=-2(\beta,\alpha,\gamma)$}}

 \put(2.75,-1){\line(0,1){21}}
\put(1,0){\line(1,0){18}}

\put(3.65,-1){\line(0,1){19}}
\put(6.7,-1){\line(0,1){18}}
\put(1,2){\line(1,0){18}}

\put(3.7,-2){\line(1,2){11.3}}
\put(12,14){\tiny $(\gamma+4)=2\beta$}

 \multiput(0.7,-0.1)(1,0){18}{\line(0,1){0.2}}
 \multiput(2.6,-0.98)(0,0.98){22}{\line(1,0){0.2}}

\put(19,2.5){\tiny $s\ell(n+1)=A_n$}
\put(19.8,1.9){\tiny $(\beta \leftrightarrow \gamma)$}

\put(3.4,18.2){\tiny $sp(2n)=C_n$}
\put(7.2,17){\tiny $so(2n+1)=B_n$}
\put(7.2,16){\tiny $so(2n)=D_n$}
\put(6.55,3.8){\tiny $\bullet D_4$ $(4,4)$}
\put(7.45,5.7){\tiny $\bullet$ $F_4$ $(5,6)$}
\put(5.85,2.55){\tiny $\bullet$}
 \put(6.55,4.85){\tiny $\bullet$}
 \put(5.7,4.8){\tiny $B_4$}
\put(5.3,3.1){\tiny $G_2$}

\put(3.5,3.75){\tiny $\bullet$ $C_2$}
\put(3.5,4.75){\tiny $\bullet$ $C_3$}
\put(3.5,5.75){\tiny $\bullet$ $C_4$}
\put(6.55,5.8){\tiny $\bullet$}
\put(5.7,5.6){\tiny $D_5$}
\put(6.55,6.65){\tiny $\bullet B_5$}

\put(6.55,1.9){\tiny $\bullet$}
\put(6.8,1.4){\tiny $D_3$}
\put(6.8,2.3){\tiny $A_3$}
\put(5,2.3){\tiny $A_2$}
\put(5.55,1.9){\tiny $\bullet$}
\put(8.5,2.3){\tiny $A_5$}
\put(8.55,1.9){\tiny $\bullet$}

\put(6.55,0.9){\tiny $\bullet$}
\put(6.8,0.4){\tiny $B_2=C_2 \; (\beta \leftrightarrow \gamma)$}

\put(14.3,19.4){\tiny $\bullet$ \; $E_8$ $(12,20)$}
\put(2,19.4){\tiny 20}

\put(10.3,11.4){\tiny $\bullet$ \; $E_7$ $(8,12)$}
\put(1.9,11.5){\tiny 12}

\put(8.4,7.6){\tiny $\bullet$ \; $E_6$ $(6,8)$}
\put(2.1,7.6){\tiny 8}

{\put(2.7,20){\scriptsize $\gamma$}}
\put(2.1,3.8){\tiny 4}

\put(6.3,-0.7){\tiny 4}
\put(8.6,-0.7){\tiny 6}
\put(14.4,-0.8){\tiny 12}
\put(19.2,0){\scriptsize $\beta$}

\end{picture}

  \vspace{1cm}

\noindent
When the  condition (\ref{exline}) is fulfilled,
the dimension (\ref{unidim02c}),(\ref{dimy2g})  of
the space of the representation  $Y_2(\gamma)$ is zero
in view of the factor $(3 \gamma - 2 t)$
in the numerator of (\ref{unidim02c}).
So the corresponding projector $\proj_{(-\frac{\gamma}{2t})}$ on this space
is also equal to zero and the parameter
$ -\gamma/(2t) $  cannot be an eigenvalue of $\hC_{\ad}$. In this case,
in the general characteristic identity (\ref{chvog}) for the operator
$ \hC_{\ad} = (\ad \otimes \ad) (\hC)$,
 the last factor $ (\hC_{\ad} + \frac{\gamma}{2t})$ will be absent
 and the universal characteristic identity
 coincides with (\ref{chexe}):
 \be
 \lb{chvog01}
 \hC_{\ad} (\hC_{\ad} + \frac{1}{2})(\hC_{\ad} + 1)
 (\hC_{\ad} + \frac{\alpha}{2t})(\hC_{\ad} + \frac{\beta}{2t}) = 0 \; .
 \ee

As we showed in Subsection {\bf \ref{uniex}},
 identity (\ref{chvog01}) for the values of the
parameters $\alpha,\beta$ given in Table 2 and Table 3 exactly
reproduces the characteristic identities for the   split Casimir operator $\hC_{\ad}$
  in the case of the exceptional Lie algebras. Note that
   both algebras $so(8)$ and $s\ell(3)$
(for the latter one has to replace the parameters $\beta\leftrightarrow\gamma$) lie on the line (\ref{exline})
and the characteristic identities
are also given by the generic formula (\ref{chvog01}).
Indeed, for the algebra $s\ell(3)$ we
have $\frac{\gamma}{2t} = \frac{1}{2}$; therefore,
the eigenvalue $(-1/2)$ of the operator $ \hC_{\ad} $ is doubly degenerated and
one of the factors $(\hC_{\ad} +1/2)$ in (\ref{chvog}) must be omitted. Wherein,
for the algebra $so(8)$ both parameters $\frac{\beta}{2t}$  and  $\frac{\gamma}{2t}$
are equal to the critical value  $\frac{1}{3}$, which gives zero in
denominators of the expressions (\ref{unidim02b}), (\ref{dimy2b})
and (\ref{unidim02c}), (\ref{dimy2g})
for the dimensions $\dim V_{(-\frac{\beta}{2t})}$
and $\dim V_{(-\frac{\gamma}{2t})}$ of the
representations $Y_2(\beta)$ and $Y_2(\gamma)$. However, these
zeros are cancelled with zeros coming from the terms
$(3\beta - 2\, t)$ and $(3\gamma - 2\, t)$ in the numerators
of the  expressions for $\dim V_{(-\frac{\beta}{2t})}$, $\dim V_{(-\frac{\gamma}{2t})}$ and these dimensions
 turn out to be $35$.
 Since the eigenvalue $-\frac{\beta}{2t}=-\frac{\gamma}{2t}=-\frac{1}{3}$
 of the operator $\hC_{\ad}$ is doubly degenerated, we must omit
one of the factors $(\hC_{\ad} +1/3)$ in (\ref{chvog})
and this identity is transformed into identity (\ref{chvog01}).

The antisymmetric part $\bP^{(ad)}_-(\ad \otimes \ad)$ decomposes
for all simple Lie algebras into a direct sum of two terms
${\sf X}_1$ and ${\sf X}_2$ (see Section {\bf \ref{hCad}}),
one of which ${\sf X}_1$ is the adjoint
representation $\ad$ with the value
of the quadratic Casimir $C_{(2)}^{(\ad)} = 1 $,  and the other representation ${\sf X}_2$ has
the value of the quadratic Casimir $C_{(2)}^{({\sf X}_2)} = 2$.
The representation ${\sf X}_2$ is reducible for the case of algebras
 $s\ell(N)$
 and irreducible for all other simple Lie algebras.
The dimension of the representations
${\sf X}_1,{\sf X}_2$ and the corresponding eigenvalues
$ \hC^{(\ad)} $ and
$\hC^{({\sf X}_2)}$  of the split Casimir operator are equal to
(cf. (\ref{XX123}))
 $$
 \begin{array}{c}
\dim {\sf X}_1= \dim \mathfrak{g} , \;\;\;\;\;\;
 \hC^{(\ad)} = -1/2 \; ,  \\ [0.3cm]
\dim {\sf X}_2=\frac{1}{2}\dim \mathfrak{g}\; (\dim \mathfrak{g}-3), \;\;\;\;\;\;
\hC^{({\sf X}_2)} = 0 \; .
\end{array}
$$
The values $\hC^{(\ad)}$ and
$\hC^{({\sf X}_2)}$
agree with the characteristic identity (\ref{idCC})
for the antisymmetrized part of $\hC_{-}$, which
is valid for all simple Lie algebras.

\section{Conclusion}
In this paper, we demonstrate the usefulness of the $\mathfrak{g}$-invariant split Casimir operator $\hC$ in the representation theory of Lie algebras. Namely, for all simple Lie algebras $\mathfrak{g}$, explicit formulas are found for invariant projectors onto irreducible representations that appear in the expansion of the tensor product  $T^{\otimes 2}$ of two adjoint representations. These
projectors are constructed in terms of the operator $\hC$. The key role in this
approach plays the characteristic identities for the split Casimir operator.
It is quite remarkable that these identities have the generic form (\ref{chvog})
which depend on Vogel parameters only. If for some algebras 
one of the factors in (\ref{chvog}) matches to a zero projector, 
then, this factor has to be omitted and the corresponding identity has
less degree in terms of the split Casimir operator.

One may hope the proposed approach will be also useful for the analysis of the structure of the tensor product of  $T^{\otimes 3}$ of three adjoint representations. The dimensions of the irreducible representations that appear in
the expansion of the tensor product  $T^{\otimes 3}$ are well known \cite{Fog},
while the structure of the  projectors is missing. We are planning to report the corresponding results elsewhere.

\section*{Acknowledgments}
The authors are thankful to P.~Cvitanovi\'{c}, R.L.Mkrtchyan,
 O.V.~Ogievetsky and M.A.Vasiliev for useful comments.
 A.P.I. acknowledges the support of the Russian Science Foundation, grant No. 19-11-00131.
 S.O.K. acknowledges the support of the Russian Foundation for Basic Research,
grant No. 20-52-12003.

\bibliographystyle{abbrv}

\end{document}